# Ultrasensitive 3D Aerosol-Jet-Printed Perovskite X-Ray Photodetector

*Anastasiia Glushkova[1], Pavao Andričević*[1], Rita Smajda[2], Bálint Náfrádi[1], Márton Kollár[1], Veljko Djokić[1], Alla Arakcheeva[1], László Forró[1], Raphael Pugin[2] and Endre Horváth[1,2]*

1Laboratory of Physics of Complex Matter (LPMC), Ecole Polytechnique Fédérale de Lausanne, Centre Est, Station 3, CH-1015 Lausanne, Switzerland;

2Centre Suisse d'Electronique et de Microtechnique (CSEM SA), CH-2002 Neuchatel, Switzerland;

ABSTRACT

X-ray photon detection is important for a wide range of applications. The highest demand, however, comes from medical imaging, which requires cost-effective, high-resolution detectors operating at low photon flux, therefore stimulating the search for novel materials and new approaches. Recently, hybrid halide perovskite $CH_3NH_3PbI_3$ ($MAPbI_3$) has attracted considerable attention due to its advantageous optoelectronic properties and low fabrication costs. The presence of heavy atoms, providing a high scattering cross-section for photons, makes this material a perfect candidate for X-ray detection. Despite the already-successful demonstrations of efficiency in detection, its integration into standard microelectronics fabrication processes is still pending. Here, we demonstrate a promising method for building X-ray detector units by 3D aerosol jet printing with a record sensitivity of 2.2 x $10^8$ µC $Gy_{air}^{-1}cm^{-2}$ when detecting 8 keV photons at dose-rates below 1 µGy/s (detection limit 0.12 µGy/s), a four-fold improvement on the best-in-class devices. An introduction of $MAPbI_3$-based detection into medical imaging would significantly reduce health hazards related to the strongly ionizing X-rays photons.

INTRODUCTION

Medical imaging remains a hallmark of modern medicine, enabling the detection and characterization of cancers, cardiovascular diseases, osteoporosis, neurological diseases and many others. In this exciting market, Computed Tomography (CT) remains highly popular as it is cheap, fast and robust in comparison to MRI.[1–3] However, despite recent hardware improvements, exposure of the human body to strongly ionizing radiation, which itself is an important health hazard, remains a major concern and a crucial limitation of CT's utility. The medical imaging industry is seeking new ways to reduce a radiation dose in CT as a key opportunity for growth. A typical CT scan exposes the patient to 2-16 mGy, which is the equivalent of one to eight years of background radiation.[4,5] It is estimated that such radiation levels contribute to 1-3% of cancers[6] due to mutations induced by X-ray photons, hence the strong motivation to search for novel, more efficient photosensitive materials which can retain good image quality at a much lower photon flux.

Numerous studies have reported the use of lead halide perovskites, $MAPbX_3$ (MA = $CH_3NH_3^+$, X = $Cl^-$, $Br^-$, $I^-$), for highly functional light-harvesting.[7] They offer promising applications in many fields, for example, the fabrication of devices such as solar cells,[8] light-emitting diodes,[9] electrochemical cells,[10] gas sensors,[11] and photodetectors.[12] The greatest success is in solar cells, where, under laboratory conditions, the certified power conversion efficiency has reached 25%.[13] This is partially attributed to the material's high light absorption coefficient, long carrier lifetime, and high charge mobility. For the above applications, the material is typically synthesized, either as single crystals by room temperature solution crystallization[14] or as polycrystalline thin films by one- and two-step deposition.[15]

Thanks to the high atomic numbers of Pb, I and Br, which results in the suitable stopping power of the material,[16,17] lead halide perovskites are also used for X-ray detection. Moreover, direct conversion of X-ray radiation into an electrical signal (photocurrent) enables higher spatial resolution compared to the competing scintillator-based technology,[18] it also comes at a reduced cost. Additionally, it has been shown that MAPbI$_3$ exhibits long-term stability even to high radiation doses,[19,20] making it a promising material for operationally-stable X-ray detectors.

Directly using the photocurrent generated by X-rays for imaging purposes, it is necessary to define a dense array of MAPbI$_3$ detector elements. The capability to deposit this photovoltaic material on various substrates with good spatial control is therefore essential for high-resolution imaging. Attempts to make arrays of 80-µm-wide wire-shaped MAPbI$_3$ by using a large-scale roll-to-roll micro-gravure printing technique[21] and to grow vertical MAPbI$_3$ nanopillars with a nanoimprinting crystallization technique[22] have been reported.

The most promising approach has been elaborated by Spina *et al.*,[23] in which nanowires of MAPbI$_3$ were grown in lithographically defined channels, so all the geometrical parameters could be precisely controlled. The appealing feature of this technique is that the design and the surface area of the nanowire array are in principle unlimited. Furthermore, to improve the responsivity of this photodetector, graphene was used to amplify the electrical signal. Due to its band structure in the form of a Dirac cone, the photoelectrons generated in MAPbI$_3$ diffuse into the graphene, creating an amplified output. Responsivity as high as 6 x 10$^6$ A/W was measured, allowing potentially a single photon detection.[12] This observation put forward the idea, that already with a modest responsivity, one could fabricate an X-ray detector with high resolution and high sensitivity, but with a strongly-reduced X-ray flux compared to the currently-used scanners.

Following this idea, the costly and relatively slow electron-beam lithography could be replaced by direct pattern writing. One of the most-used direct writing methods is ink-jet printing, which has enabled a variety of applications.[24–26] Recent developments in this field have led to low-cost production of sensing electronics.[27] Many functional materials, including perovskites,[28] can be deposited with such a method by varying the composition of the ink.[29] Wei *et al.* developed the two-step process to pattern MAPbI$_3$ by ink-jet printing MAI on a spin-coated lead iodide (PbI$_2$) layer.[29] However, this might impose technical constraints since PbI$_2$ has to be coated onto the whole surface of the designed element.

Our approach for MAPbX$_3$ integration into electric circuits is based on aerosol jet printing (AJP). With AJP, the material can be selectively deposited at any location of interest with μm precision. The success of this technique in writing well-defined 3D structures is linked to the existence of intermediate phases of MAPbX$_3$ formed with polar aprotic solvents in the form of elongated crystallites. These solvatomorph phases are pre-formed in the nozzle, grow during the time-of-flight, and land on the substrate as already-formed crystalline nanowires containing little solvent. This prevents the spreading of the solution on the surface, and/or the dissolving of the underlying layers in the case of defining high-aspect-ratio 3D structures. Once at the surface, with further evaporation of the solvent, the intermediate phases transform into MAPbX$_3$. This deposition feature is very specific to MAPbX$_3$ chemistry with polar aprotic solvents and allows the designing of a well-defined network, necessary for an X-ray photodetector. Here we demonstrate the fabrication and characterization of a unit consisting of MAPbI$_3$ on graphene. Our device shows ultra-high sensitivity of 2.2 x 10$^8$ µC Gy$_{air}^{-1}$ cm$^{-2}$, for detecting 8 keV X-rays photons at low radiation dose-rates.

RESULTS AND DISCUSSION

**MAPbX$_3$ aerosol jet printing.** AJP enables the precise patterning, deposition, and unlimited layering of a functional material, with the possibility to create arrays or other configurations of elements on various substrates. During the ink preparation, MAPbI$_3$ and MAPbBr$_3$ crystals are stabilized as soluble complexes in dimethylformamide (DMF). Solutions of MAPbI$_3$ and MAPbBr$_3$ (see Methods) in the polar aprotic solvent are loaded in the reservoir, as depicted in Fig. 1a. Using a flux of nitrogen gas, the generated aerosol is transported to the nozzle in the form of micro-droplets and focused at its tip (details on parameters of the process are described in Methods and Supplementary (SI) Fig. S1). During the flight, the solvent starts to evaporate, leading to the homogenous nucleation and crystallization of perovskite intermediate phases already in the nozzle. When the droplet reaches the substrate, the solvent is mostly evaporated, thus reducing the spatter of the solution and improving spatial precision. Moreover, preventing dissolution of the previously-deposited layers allows the creation of 3D patterns, for example, high aspect ratio pillars (Fig. S2), layer by layer, as illustrated in Fig. 1b-g. Likewise, an average optimized spot size of 48.7 ± 2.8 µm is attained, while the previously-achieved minimal spot size from MAPbI$_3$ ink was 75 µm, demonstrated by Mathies *et al.*[30] In AJP MAPbI$_3$ the number of grain boundaries is large, presumably resulting in the increase in charge trapping. This leads to the trap-induced photoconducting gain, which helps in obtaining higher detector sensitivity.[31]

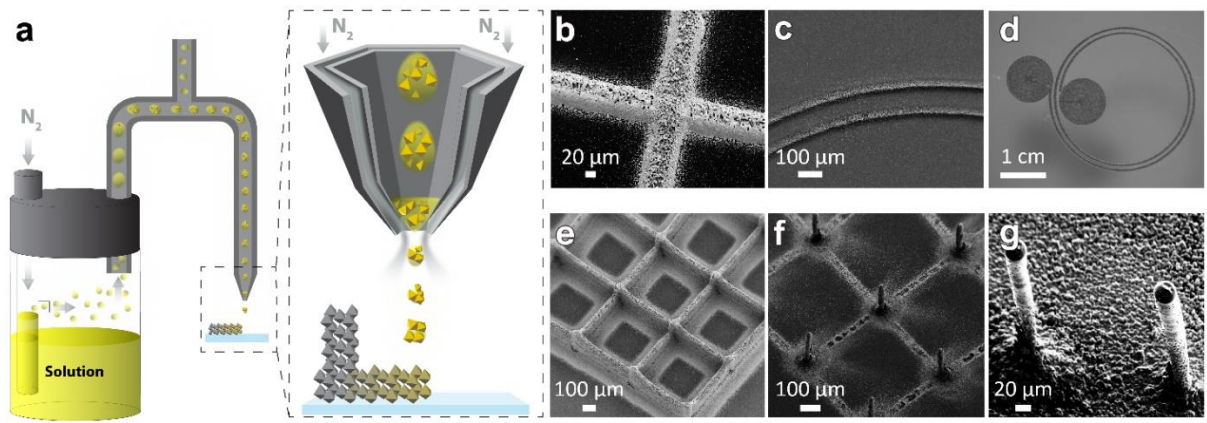

**Figure 1.** 3D MAPbX$_3$ aerosol jet printing. (a) Schematic of the AJP system. The jet-focusing nitrogen flow helps the fast evaporation of the solvent and the growth of the intermediate phases (see main text), which is important for the creation of well-defined 3D structures. (b-g) Various printed patterns of MAPbI$_3$ lines, spirals, grids and pillars written on glass substrate.

**MAPbX$_3$ crystallization under fast solvent evaporation conditions.** Since the homogenous nucleation and crystal growth of MAPbX$_3$ intermediates under fast solvent evaporation in the flying micro-droplets will have an impact on crystal quality and crystal morphology of the final MAPbX$_3$ material,[32–34] it is important to understand the sequence of phase transformations. In order to mimic this intermediate phase crystallization process in the AJP typically occurring in sub-second flight time at a slower speed, we recorded X-ray diffraction (XRD) data at the ESRF synchrotron radiation source during the crystallization process with a time-lapse of 1 s. Thanks to this high-resolution technique, we could detect the growth of the first observable nuclei (SI Fig. S3) and the following dynamics of MAPbI$_3$ crystallization. The XRD profiles are presented in Fig. 2a. The phase analysis based on the Rietveld refinements of these complex patterns gives two intermediate phases prior to the MAPbI$_3$ formation. These are (MA)(DMF)PbI$_3$ and (MA)$_2$(DMF)$_2$Pb$_3$I$_8$ (details of the analysis are given in Supplementary (SI) Fig. S4-6). A strongly-dominant Bragg reflection at 2θ = 3.7 °, observed at the first stage of the crystallization of the (MA)(DMF)PbI$_3$, corresponds to the structure planes (011) (SI Fig. S4b) which indicates the [011] direction of the elongation of the needle-shaped crystallites, perpendicular to this plane (as shown in Fig. 2b). During the experiment, this reflection changed its width over time. Using the Scherrer equation,[35] one can deduce the mean size of the crystallites in the direction of their elongation. From 63 nm at the beginning of the crystallization, it evolves to a maximum size of 78 nm before declining for the profit of MAPbI$_3$. The estimation of the normalized intensity and the mean size is depicted in the SI Fig. S4c and the time-dependent evolution of the volume fractions in the SI Fig. S6. One has to notice that the sample temperature should be raised above room temperature in order to evaporate the DMF solvent and to transform (MA)(DMF)PbI$_3$ and (MA)$_2$(DMF)$_2$Pb$_3$I$_8$ into MAPbI$_3$. In this XRD experiment, a temperature of 64°C was necessary for the transformation of (MA)(DMF)PbI$_3$, while the remaining (MA)$_2$(DMF)$_2$Pb$_3$I$_8$ is present in about 13vol% (see SI Fig. S4a and S6). This confirms that within this time scale, for a complete phase transformation to MAPbI$_3$, a minimum 70 °C is required, in agreement with earlier observations for this phase.[31]

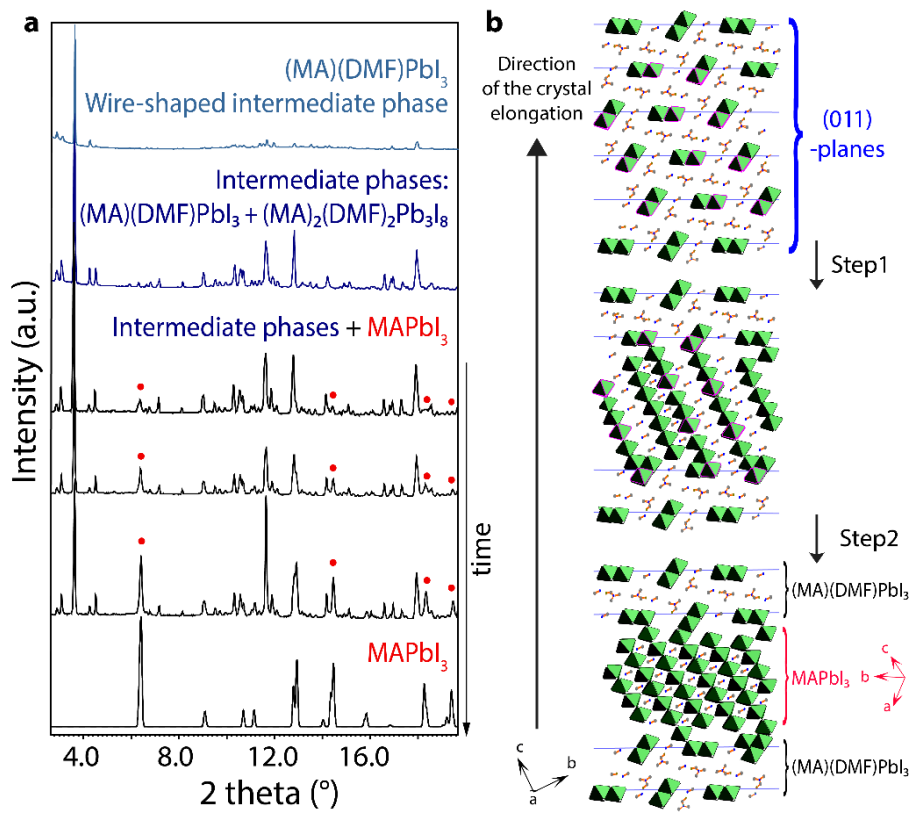

**Figure 2.** Crystallization of MAPbI$_3$ from intermediate solvatomorph phases. (a) The experimental X-ray diffraction patterns recorded during crystallization: from formation of the intermediate phases (MA)(DMF)PbI$_3$, (MA)$_2$(DMF)$_2$Pb$_3$I$_8$ to MAPbI$_3$. The reflections related to MAPbI$_3$ and none of the intermediate phases are indicated by red dots. (b) Modelling of the MAPbI$_3$ formation from (MA)(DMF)PbI$_3$ (sp. group P2$_1$/c; a = 4.5647(9), b = 25.446(5), c = 12.119(2) Å, β = 96.75(3) °) based on the structural studies.[34] The wire-like MAPbI$_3$ grows in the direction perpendicular to the (011) planes of (MA)(DMF)PbI$_3$.

We put forward two possible ways for this phase transformation. Depending on the DMF diffusion kinetics, it may start at the surface or in the core of the single (MA)(DMF)PbI$_3$ crystallite. Under rapid solvent evaporation, the DMF diffusion from the core to the surface is slow, resulting in the MAPbI$_3$ formation on the surface. This leads to the formation of MAPbI$_3$ shell and an encapsulation and stabilization of (MA)(DMF)PbI$_3$ inside the core of the two-phase product (see modelling in SI Fig. S7). We observed this in the synchrotron radiation experiments when the X-ray beam accelerates the DMF evaporation from the crystalline surface. In the case of a slow solvent evaporation rate, DMF diffusion from the core to the surface initiates the transformation of a crystallite as depicted in Fig. 2b. The second scenario leads to a completed transformation and mono-phase, the MAPbI$_3$ product. Precisely this case is reached in the AJP process.

**Characterization of aerosol-jet-printed MAPbX$_3$.** The morphology and uniformity of the polycrystalline AJP-deposited MAPbX$_3$ are important for efficient X-ray detection. The morphology strongly depends on the saturation of the solution and the temperature of the substrate during the deposition process.[36] Arrays of MAPbBr$_3$ emitters with diameters of 93 µm and 48 µm have been deposited and examined by fluorescence imaging and SEM (SI Fig. S8). The homogeneous fluorescence emission intensity indicates the good optical quality and structural reproducibility of the deposits.

The photodetection properties of an AJP-deposited MAPbI$_3$ pattern were determined under visible-light illumination. MAPbI$_3$ of different geometries were printed on microstructured surfaces (see Methods). The 3D printed geometry (pillar shape) showed an increased charge carrier collection compared to its 2D counterpart (spot shape), SI Fig. S9. Both 2D and 3D geometries enabled detection of low light intensities, down to 31.4 µW/cm$^2$ under continuous illumination ($\lambda$ = 650 nm, SI Fig. S10) due to a very stable dark current.

**MAPbX$_3$/graphene heterostructures *via* 3D AJP.** To further increase charge collection of the photodetection characteristics of 3D printed halide perovskites light sensors, the strategy of Spina *et al*.[12] was partially adopted by preparing MAPbI$_3$/graphene heterostructures *via* AJP (SI Fig. S11). Graphene has a low light absorption of 2.3 %,[37] so by itself is not ideal for photodetection. But due to its band structure, the zero band-gap character,[38] it could amplify the effect of photoelectrons created in MAPbX$_3$ in the following way. The high photogain is a result of the very high defect density near the MAPbI$_3$–graphene interface, as explained in SI Figure S12. Under illumination, photogenerated electrons are trapped at these defects, leading to lowering the Fermi level in the graphene from the Dirac level,[39] which is facilitated by the photogenerated holes. The lowered Fermi level causes a large increase in the electronic states which contribute to current conduction, resulting in a drastic increase in current through the graphene.[40] Overall the high photogain is obtained as a small amount of photogenerated e-h pairs cause orders of magnitude more carriers to pass through the graphene.

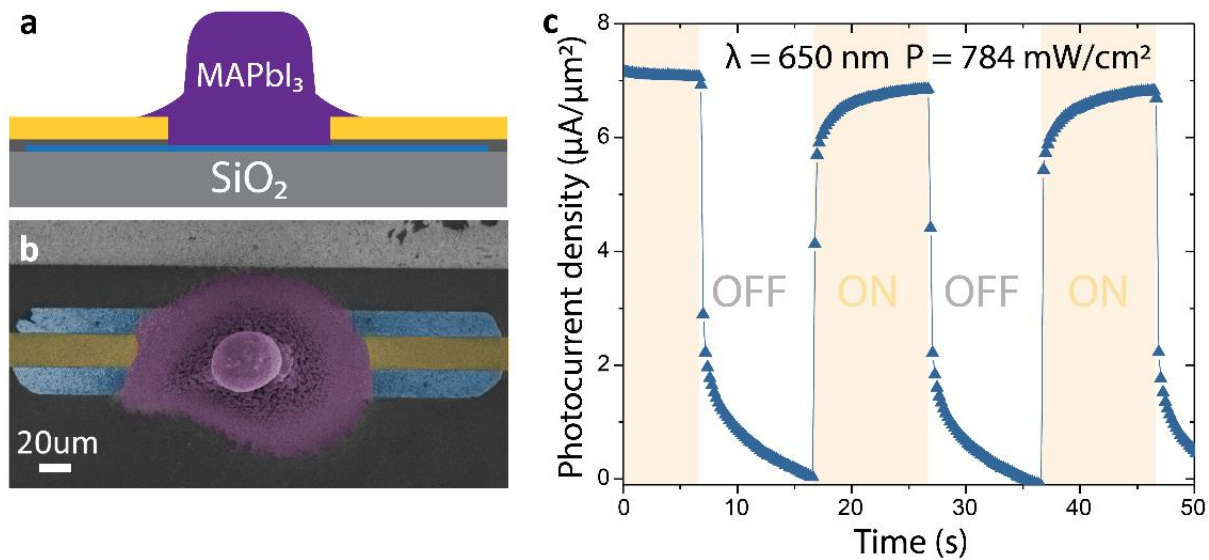

**Figure 3.** MAPbX$_3$/graphene heterostructured sensing unit for photo-detection. (a) Schematic drawing of the side view and, (b) top view of false-colored SEM images of a sensing unit, an AJP-deposited MAPbI$_3$ pillar (purple) on the graphene layer (blue) with gold electrodes (yellow). (c) Visible light-induced photocurrent responses at 10 mV bias voltage.

In order to determine the optimal pixel design and sensing configuration, several possible architectures and operation modes have been tested (resistor, diode, and heterostructure, as illustrated in SI Fig. S13). The best performance was found for MAPbI$_3$ pillars printed on a graphene layer, covering the whole distance between the gold electrodes (Fig. 3a-b). The device demonstrated good detection characteristics even for low bias voltages, making it viable for low-power electronics use (Fig. 3c). This structural unit represents one pixel in the detector. The operational stability of the AJP MAPbI$_3$-graphene pixel was tested for 30 minutes under a constant voltage and a 358 µGy/s dose-rate, Figure S14. The device showed no sign of degradation during the test. A signal drift of 1.5 µA was observed, which is probably due to ion migration, a common feature of MAPbX$_3$ materials.[41]

The graphene layer was microstructured in the form of stripes, on which the electrodes were evaporated, and in the gaps the MAPbI$_3$ pillars were 3D AJP-printed. A simplified layout was tested, where the graphene sheet was uniform, on which the microelectrodes and the perovskite materials were deposited. Even in this configuration, tested for white light and X-rays, the impact of the photon could be well localized and the cross-talk between the pixels was minimal (SI Fig. S15, S16 and S17). Therefore, this simpler layout still gives a good image contrast.

**MAPbX$_3$/graphene heterostructured X-ray photodetector.** Because the photon absorption conditions for high energetic X-rays are different than for visible light,[19] their impact in the smallest MAPbI$_3$ spot size gives a weaker response (SI Fig. S18). Therefore, devices with a wall of MAPbI$_3$ of about 600 µm high were fabricated, as shown in Fig. 4a-c (see Methods). X-ray detection properties of the devices were investigated using a standard fine-focus copper X-ray tube (CuK$_\alpha$). X-ray dose-rates were selected by varying tube voltage and current in the range of 15 kV - 45 kV and 5 mA - 40 mA, respectively. The most sensitive pixels were exposed to dose-rates from 0.12 µGy/s to 358 µGy/s while opening and closing the shutter in front of the 8 keV X-ray source (Fig. 4d-e) for periods of 10 seconds. The X- ray detection limit was found to be at a 0.12 µGy/s dose- rate, below which the signal could not be distinguished from the background noise. High photocurrents of up to 4000 µA/cm$^2$ were measured at 358 µGy/s dose-rate (Fig. 4d).

The sensitivity, $S$, of the device is dose-rate dependent. The reason for this effect is that at higher dose-rates, photo-induced holes transferred into the graphene layer and their corresponding electrons left in the perovskite layer create an additional electric field weakening the original field near the electrode interface. This leads to a decrease in the photocurrent, ultimately resulting in a lower device sensitivity at higher dose-rates.[42] Accordingly, we calculated $S$ at 100 mV bias for three regions by linear fitting $I_{ph}=I_d+S\cdot D$ to the experimental data, where $I_{ph}$ is the observed photocurrent, $I_d$ is the dark current, $S$ is the sensitivity and $D$ is the X-ray dose-rate. At dose-rates below 1 µGy/s, the sensitivity of the device is the greatest, reaching $S=2.2 \times 10^8$ µC $Gy_{air}^{-1}$ $cm^{-2}$. It is important to point out that this is the sensitivity of the pixel on the chip with an estimated surface of 0.075 $mm^2$. For dose-rates above 1 µGy/s, the photocurrent begins to saturate (Fig. 4e), yielding a sensitivity of $S=2.5 \times 10^7$ µC Gy$^{-1}$cm$^{-2}$ for a dose-rate of up to 40 µGy/s and finally, $S=2.9 \times 10^6$ µC Gy$^{-1}$cm$^{-2}$ was found from 40 to 100 µGy/s. A linear dependence of the response was measured (SI Fig. S19).

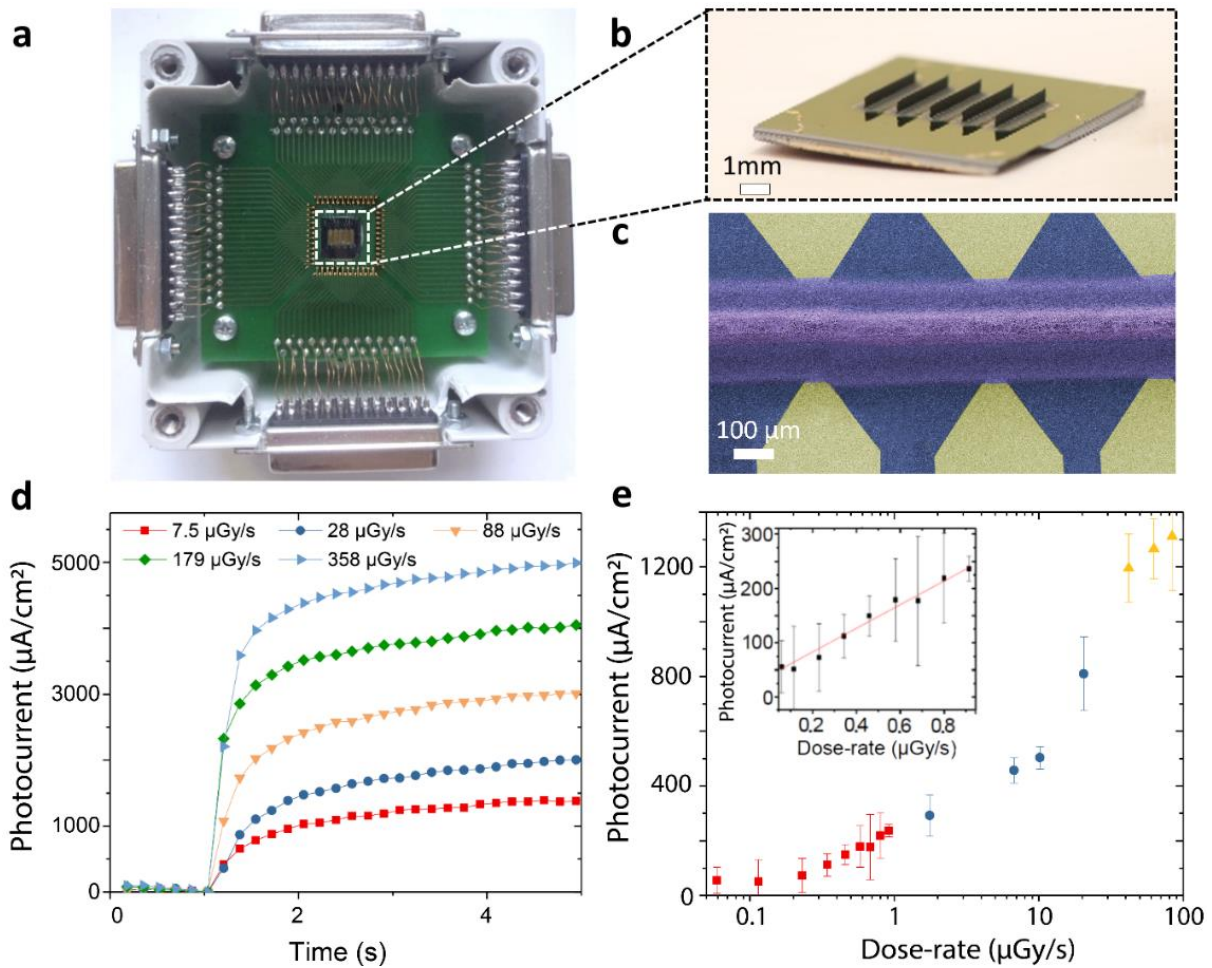

**Figure 4.** X-ray detector measurements. (a) Photograph of the fully integrated X-ray detector. (b) 1 cm2 sensing chip with 3D printed $MAPbI_3$ walls about 600 µm height. (c) False-colored SEM image of the 3D printed $MAPbI_3$ wall on the Ti/Au electrodes (graphene in blue, $MAPbI_3$ in purple and metal electrodes in yellow). (d) X-ray illumination-induced photocurrent response as a function of time at 100 mV bias voltage. (e) Photocurrent density as a function of X-ray dose-rate, region below 1 µGy/s is in the inset.

One of the major obstacles preventing perovskite-based optoelectronics from being commercialized is its stability. The active layer in the device, MAPbI$_3$, should be protected from water vapour present in ambient air. This also applies to the X-ray detector unit. After the wire-bonding (Fig 4a), it was encapsulated with a PDMS polymer, which seals the material well. The shelf-life stability was tested and the operational reproducibility checked monthly after assembling the detector unit. It has shown satisfactory stability since, in the time period of nine months, no degradation in performance was observed (SI Fig. S20).

CONCLUSION

Using a rapid and cheap printing technique, the AJP, we have structured and tested an X-ray detector unit, which shows a record sensitivity of $2.2 \times 10^8$ µC Gy$_{air}^{-1}$cm$^{-2}$, four orders of magnitude higher than the up-to-date perovskite devices. Furthermore, MAPbX$_3$-based devices have recently experienced exceptional radiation hardness to (no degradation under the impact of even high energy photons, > 1 MeV). It is postulated that this is due to a self-healing mechanism happening at the nanoscale.[43] Altogether, we are confident that this X-ray detector unit architecture is extremely promising for high sensitive X-ray imaging. A custom-designed X-ray detector will consist of assembling such units into a large surface area. This approach could allow significant lowering of the radiation doses required for X-ray imaging, resulting in safer and more affordable Computed Tomography imaging systems. However, it is important to point out that medical devices, such as CT scans, use higher energies of X-rays. Therefore, detection measurements should be repeated for photons in the 100 keV range to confirm the device capability to be utilized for medical applications

EXPERIMENTAL SECTION

*Solution preparation*. All of the purchased chemical reagents were of analytical reagent grade purity and used without further purification. The perovskite precursor $PbI_2$ and MAI/MABr (molar ratio 1:1) dissolved in dimethylformamide (DMF) with a concentration of 15 wt% for $MAPbI_3$ and $MAPbBr_3$ were used for aerosol jet printing.

*3D printing.* Aerosol jet printing was performed using a commercial Aerosol Jet Printer, Optomec AJ-300. Printed patterns are created by translating the deposition head with respect to the substrate in XY directions using a path generated from a CAD file. AJP requires control over numerous parameters, including the flow rate of the carrier and sheath gases, temperature and speed of the stage. A vial was loaded with 10 ml of the ink (solution). The distance between the nozzle tip (150 μm diameter) and the substrate was kept constant at 0.92 mm. The speed stage was 1 mm/s. The sheath gas flow was fixed at 693.7 sccm, while the carrier gas flow was varied. Figure S1a shows the AJP $MAPbI_3$ at a carrier gas flow rate of 29.4 sccm and stage temperature at 19.6 °C. With this conditions the intermediate phases were not formed during the flight and the final material did not complete the transformation to $MAPbI_3$. Figure S1b demonstrates the pillar growth with optimized conditions for the carrier gas flow rate (59.4 sccm) and stage temperature (45°C).

During the nozzle multipass, the spread of the material around the 3D feature grew. This results in a total spread width of 250 μm of a 45 μm patterned line. For this height, 40 repetitions of the patterning were made. And after 120 repetitions the spread width was 585 μm.

*Device fabrication.* The AJP spot and pillar characteristic comparison were performed on the samples on the 300 nm $SiO_2$/Si substrate with Ti-Au (20 - 50 nm) electrodes. For these samples, electron beam lithography (MMA/PMMA) and a lift-off process were carried out at CMi (Center of MicroNanoTechnology), EPFL. The gap between the electrodes varied from 60 μm to 140 μm. Different configurations of the heterostructure were fabricated on graphene purchased from commercial vendors (Graphenea). The monolayer graphene was produced by CVD and transferred to a substrate of $SiO_2$/Si (300 nm dry oxide) by a wet transfer process. The pattern for the diode-like structure and non-etched samples were exposed with an electron beam lithography. After the first electron beam lithography, the graphene layer was ion-beam etched; with the second electron beam exposure the electrodes were patterned, and Ti-Pt (20 - 70 nm) was evaporated. After the lift-off process, the sample was ready to be printed on. For the X-ray detector, the Ti-Au (20 - 50 nm) electrodes were evaporated, as described above, omitting the first step of graphene etching.

*XRD measurements and analysis.* Several microliters of $MAPbI_3 \cdot DMF$ solution (30 wt%) were pipetted on the fused quartz capillary and subjected to the X-ray beam at room temperature. The synchrotron radiation powder X-ray diffraction experiments were carried out under Oxford Cryostream 700+ $N_2$ blower with $\lambda = 0.71446$ Å for an exposure time of 0.9977 s using the PILATUS@SNBL detector at the Swiss–Norwegian Beam Lines, European Synchrotron Radiation Facility[43]. The powder diffraction data were processed with BUBBLE software.[44] When the intermediate phases were stabilized at room temperature (after 24 minutes), the sample was heated to 64 °C to evaporate the trapped solvent. The absence of long exposure damaged has been checked by probing a fresh spot on the sample. The Crys-AlisPro (Oxford Diffraction) program package was used for the experimental data processing. Structural calculations were made with JANA2006 software.[45]

*Device characterization.* The optimization of the device configuration (three types - resistor, diode, and heterojunction) was performed using a Keithley 220 current source and a Nanovoltmeter 2182 and Keithley 2400 source meter. The photocurrent was measured in ambient conditions. Two-point measurements were performed under white monochromatic light (450 nm, 550 nm, 650 nm) with a sweep in intensities from 293.71 nW to 0.48 nW. The X-ray-generated photocurrent was measured in a two-terminal geometry under fixed DC-bias. X-ray detection properties of the devices were investigated using a standard fine-focus copper X-ray tube (CuK$_\alpha$; $\lambda$ = 1.541874 Å, take-off angle: 6.0° - for the first measurement) mounted in a PANalytical Empyrean diffractometer, energized from 15 kV to 45 kV and 5 mA to 40 mA. A commercial diffractometer detector (FLUKE Biomedical - VICTOREEN® Advanced survey meter 990S) was used for calibration of the X-ray intensity (more information on the calibration is in Supporting Information, Figure SI 20-21). A Keithley 2400 source meter allowed us to measure the current with less than 0.1 nA resolution while tuning the applied bias voltage, in the dark and under X-ray irradiation. For low intensity, a 0.2 mm Cu plate was used to attenuate the beam to reach a 0.12 µGy/s dose-rate. All performance measurements of the devices were carried out in ambient conditions at room temperature.

ASSOCIATED CONTENT

The following files are available free of charge.

Additional information on the nucleation and early stage of crystal growth of MAPbI$_3$ during the 3D printing process; different phase transitions tracked *via* synchrotron XRD. SEM images of different device designs and a comparison of their performance. Operational stability and signal cross talk testing. Detailed characterization of the best performing pixel device during from fabrication until submission. Table comparing our device with the best in class perovskite based X-ray detectors.

## AUTHOR INFORMATION

**Corresponding Author**

Email: pavao.andricevic@epfl.ch

**Author Contributions**

The manuscript was written through contributions of all authors. All authors have given approval to the final version of the manuscript. E.H. and R.S. conceived the study, E.H. developed the solution preparation method. E.H. and R.S. carried out the deposition and photoluminescence measurements. E.H. and M.K. prepared the halide perovskite solutions. B.N. designed the optical and X-ray photosensitivity measurements. A.G. performed SEM imaging and fabricated the chips for deposition and integrated them for measurements. A.G., P.A., V.D. carried out the photocurrent measurements under visible light and X-ray irradiation. A.G., P.A. performed the data analysis. A.A. and A.G. carried out XRD experiments. A.A. refined and analyzed XRD data. L.F. supervised the project. All authors discussed the results and assisted in the preparation of the manuscript.

## ACKNOWLEDGMENT

The authors are grateful for the financial support of the ERC Advanced grant Picoprop (Project Number: 670918) and PoC ERC grant Picoprop4CT. The authors wish to acknowledge BM01 staff for the support and help with the XRD experiment. Allocation of the beamtime at the Swiss–Norwegian Beam Lines (SNBL) by SNX council is greatly appreciated. We thank the

TABLE OF CONTENTS IMAGE (TOC)

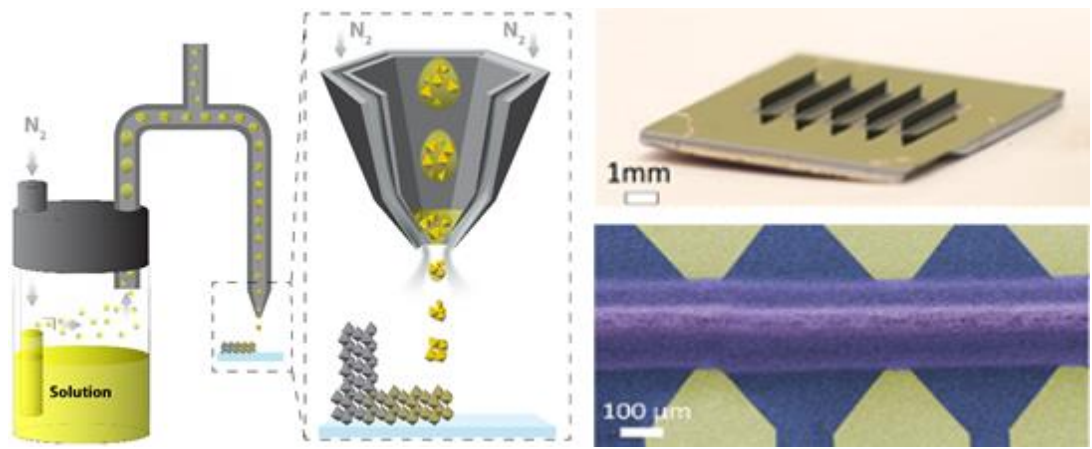